\documentclass[conference]{IEEEtran}
\IEEEoverridecommandlockouts
\usepackage{cite}
\usepackage{amsmath,amssymb,amsfonts}
\usepackage{algorithmic}
\usepackage{graphicx}
\usepackage{textcomp}
\usepackage{xcolor}
\usepackage{xspace}
\usepackage{float}
\usepackage{amssymb, amsmath}
\usepackage[T1]{fontenc}
\usepackage{cite}
\usepackage{orcidlink}
\usepackage{academicons}
\usepackage{tabularx}
\usepackage{booktabs}
\usepackage{adjustbox}
\usepackage{nicematrix}
\usepackage{multirow}

\def\BibTeX{{\rm B\kern-.05em{\sc i\kern-.025em b}\kern-.08em
    T\kern-.1667em\lower.7ex\hbox{E}\kern-.125emX}}
\begin{document}

\title{M$^2$UNet: MetaFormer Multi-scale Upsampling Network for Polyp Segmentation
}

\author{
\IEEEauthorblockN{ Quoc-Huy Trinh\orcidlink{0000-0002-7205-3211}, 
Nhat-Tan Bui \orcidlink{0000-0002-4303-1582}, 
Trong-Hieu Nguyen-Mau\orcidlink{0000-0003-2823-3861},  \\
Minh-Van Nguyen\orcidlink{0009-0009-0309-654X},
Hai-Minh Phan\orcidlink{0009-0005-9570-8689}, 
Minh-Triet Tran\orcidlink{0000-0003-3046-3041}, Hai-Dang Nguyen\orcidlink{0000-0003-0888-8908}}
\IEEEauthorblockA{
\textit{Faculty of Information Technology, Software Engineering Laboratory,}
}
\IEEEauthorblockA{
\textit{International Training and Education Center, and John von Neumann Institute}
}
\IEEEauthorblockA{
\textit{University of Science, VNU-HCM}
}
\IEEEauthorblockA{
\textit{Vietnam National University, Ho Chi Minh City, Vietnam} 
}

20120013@student.hcmus.edu.vn, 
1859043@itec.hcmus.edu.vn, 
20120081@student.hcmus.edu.vn,\\
nmvan20@clc.fitus.edu.vn,
phminh22@clc.fitus.edu.vn,
tmtriet@fit.hcmus.edu.vn, nhdang@selab.hcmus.edu.vn
\vspace{-1mm}
}


\maketitle


\begin{abstract}
Polyp segmentation has recently garnered significant attention, and multiple methods have been formulated to achieve commendable outcomes. However, these techniques often confront difficulty when working with the complex polyp foreground and their surrounding regions because of the nature of convolution operation. Besides, most existing methods forget to exploit the potential information from multiple decoder stages. To address this challenge, we suggest combining MetaFormer, introduced as a baseline for integrating CNN and Transformer, with UNet framework and incorporating our Multi-scale Upsampling block (MU). This simple module makes it possible to combine multi-level information by exploring multiple receptive field paths of the shallow decoder stage and then adding with the higher stage to aggregate better feature representation, which is essential in medical image segmentation. Taken all together, we propose MetaFormer Multi-scale Upsampling Network (M$^2$UNet) for the polyp segmentation task. Extensive experiments on five benchmark datasets demonstrate that our method achieved competitive performance compared with several previous methods.
\end{abstract}


\section{Introduction}


Colorectal cancer poses a significant threat to human health and society, making it a substantial health concern. Polyps, which are abnormal growths in the colon or rectum, have the potential to develop into cancerous tumors over time. Early detection of polyps plays a vital role in preventive healthcare, as it can greatly improve the prognosis and treatment effectiveness for individuals with colorectal cancer \cite{colectoral}.


In recent years, early diagnosis has emerged as a critical factor in treating and preventing colorectal cancer, particularly in polyp detection. However, the accuracy of early diagnosis is constrained by various external factors, as highlighted in \cite{lack-endoscopic}. Consequently, polyp segmentation has become an integral component of the diagnostic process. 

Deep Learning approaches have gained prominence in polyp image segmentation, with methods such as UNet \cite{unet}, UNet++ \cite{unet++}, PraNet \cite{pranet}, MSNet \cite{MSNet}, and PEFNet \cite{mmm2023}, demonstrating competitive performance in state-of-the-art results. While Convolutional Neural Network (CNN) models excel in capturing local information, they face challenges in comprehensively representing the overall shape of polyp objects, which is crucial for precise segmentation. This limitation significantly contributes to missed segment areas, which are essential outputs in segmentation tasks. Moreover, existing methods mainly concentrate on improving the feature representations of the encoder and skip connection but forget to consider the decoder.



In this paper, we propose the \textbf{M}etaFormer \textbf{M}ulti-scale \textbf{U}psampling Network (M$^2$UNet) that combines MetaFormer \cite{metaformer} with UNet \cite{unet} and a Multi-scale Upsampling block (MU). The MetaFormer framework \cite{metaformer} facilitates incorporating both local and global information by employing convolution-based downsampling to capture local features while utilizing a Transformer encoder to capture global features in subsequent stages.

Besides, our MU module enhances the ability to capture multi-level information between multiple decoder stages, further ameliorating the segmentation results of the entire architecture. The MU module specifically employs two receptive field paths to exploit the information of feature maps in different aspects. The output features of MU are cumulative with the features of the decoder layer one stage away, which helps the model extracts comprehensive information from different levels of the decoder. Our proposed method shows competitive results on various datasets, demonstrating our model's ability against the weaknesses of existing approaches.

To summarize, our contributions are threefold:
\begin{itemize}

 \item We propose the \textbf{M}etaFormer \textbf{M}ulti-scale \textbf{U}psampling Network, termed as M$^2$UNet, combining MetaFormer  with UNet for improving the local and global contextual representations of the polyp objects.

 \item We introduce a Multi-scale Upsampling block (MU) for enhancing the representation ability of different levels of decoder features.


 \item We demonstrate the effectiveness of our method on five benchmark datasets: Kvasir-SEG \cite{kvasir-seg}, CVC-ClinicDB \cite{clinicdb}, CVC-ColonDB \cite{colondb}, ETIS \cite{etis} and CVC-300 \cite{cvc300}.
\end{itemize}

The content of this paper is organized as follows. In Section~\ref{sec:RelatedWork}, we briefly review existing methods related to this research. Then we propose our methods in Section~\ref{sec:Methods}. Experiments and discussion are in Section~\ref{sec:Experiments}. Finally, we present the conclusion in Section~\ref{sec:Conclusion}.

\section{Related Work}
\label{sec:RelatedWork}

\begin{figure*}[t!]
    \centering
    \includegraphics[width=1\linewidth]{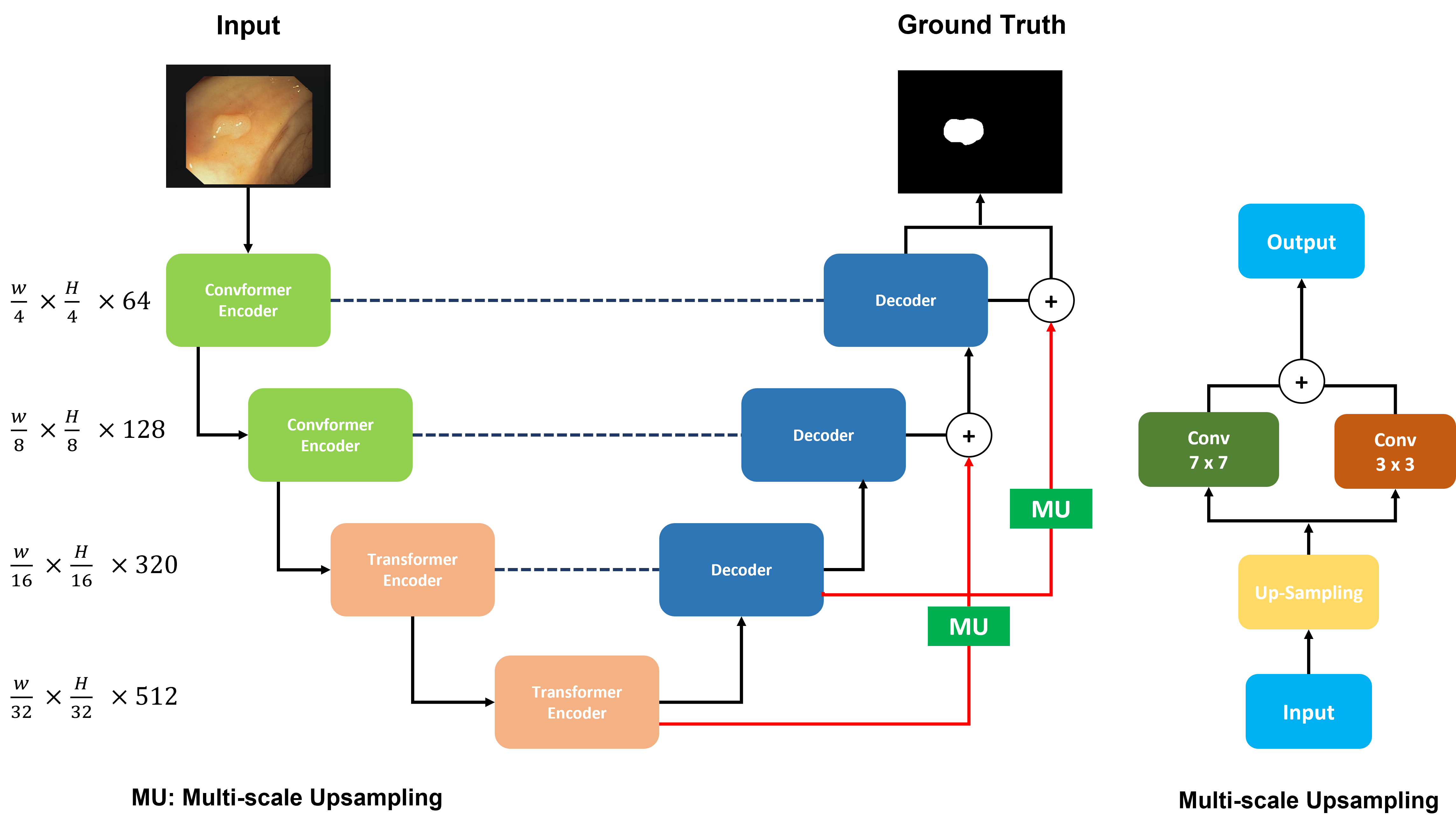}
    \caption{General architecture of M$^2$UNet}
    \label{fig:archi}
\end{figure*}


UNet \cite{unet} sets a new precedent as it was the first model to integrate skip connections in the encoder-decoder architecture, specifically for medical segmentation tasks. This pioneering technique merges shallow and deep features, improving the accuracy and reliability of the segmentation process. Since then, many works have been taken to enhance UNet's performance in the segmentation task. 

UNet++ \cite{unet++} ameliorates the performance of UNet by nested skip connections. In addition, SFA \cite{sfa} proposes the boundary-sensitive loss and additional decoder to encourage the model to focus on the polyp boundary. PraNet\cite{pranet} utilizes a reverse attention module and parallel partial decoder to enhance the precise boundary separating a polyp from the surrounding mucous membrane. MSNet\cite{MSNet} specifically employs cascaded subtraction operations at multiple levels and stages to capture complementary information from different levels. 

Recently, Hieu \textit{et al.} \cite{rivf}, and PEFNet \cite{mmm2023} concentrate on the positional information, improving the overall detection of polyp regions. In general, most methods seem to ignore useful information of the decoder's features. We believe that by capturing the multi-level information of the decoder, the model can achieve more accurate polyp localization.

In summary, UNet and its variants, along with techniques like skip connections, boundary-sensitive loss, attention mechanisms, and positional cues, have significantly advanced polyp segmentation. But there is still potential in harnessing the decoder's features to further improve accuracy.

\vspace{3mm}
\section{Methods}
\label{sec:Methods}

\subsection{General architecture}
 A novel M$^2$UNet is built by extending the UNet architecture by incorporating modifications inspired by the ConvFormer and Transformer blocks from the MetaFormer baseline in the encoder stage. Additionally, we introduce the utilization of the Multi-scale Upsampling module (MU) to enhance the multi-scale representation of the decoder. The complete architecture is illustrated in Figure~\ref{fig:archi}.

The input tensor $X \in R^{W \times H \times 3}$ has a shape of $W \times H \times 3$, where $W$ represents the width, $H$ represents the height, and the encoder extracts features $X_i \in R^{\frac{W}{2^{i+1}} \times \frac{H}{2^{i+1}} \times F_i}$ at different stages, denoted by $F_i \in \{64, 128, 320, 512\}$ and $i \in \{1, 2, 3, 4\}$, representing the filters used at each step of the encoder and decoder stages.

In the decoder stage, although the feature is upsampled twice using Convolution Transpose 2D at each step, we further enhance the feature by employing our Multi-scale Upsampling block (MU), which upsamples the feature four times at step $i$. The upsampled features are then merged with the features at step $i-2$ to enrich the feature representation while incorporating the previously upsampled features.

Following this, the decoder stage generates a mask with a shape of $W \times H \times 64$. Subsequently, a $1 \times 1$ Convolution layer is applied to map the feature map from 64 filters to a single filter, producing the final output.


\subsection{ConvFormer Block}

In the MetaFormer \cite{metaformer}, the ConvFormer block investigates the capabilities of established token mixers to achieve excellent performance. Instead of developing new token mixers, the block relies on commonly-used operators to assess the lower limits of performance and model versatility.

The ConvFormer block consists of four steps. The first step involves creating token mixers. To accomplish this, the Convolution operation is employed using two specific techniques: Depthwise Convolution and Separable Convolution. These convolutional operations are utilized to enhance the mixing of tokens within the model architecture, promoting effective information exchange and integration across different parts of the input sequence.




\begin{align}
    Conv(X_{i}) = Conv_{pw2}(Conv_{dw}(\sigma(Conv_{pw1}(X_{i}))))
    \label{equa:1} \\
    X_{i} = X_{i} + Conv(Norm(X_{i}))
    \label{equa:2} \\
    X{i} = X{i} + \sigma(Norm(X{i})W_{1})W_{2}
    \label{equa:3}
\end{align}

Equations \ref{equa:1} \ref{equa:2}, and \ref{equa:3}, referenced in the paper \cite{metaformer}, introduce the ConvFormer block. Within this equation, $Conv_{pw}$ represents the pointwise Convolution operation applied at index $i$, while $Conv_{dw}$ refers to the Depth-wise Convolution operation. In the subsequent stages, the output of equation \ref{equa:1} is normalized before the skip connection is incorporated. Following the normalization step, equation \ref{equa:2} demonstrates the application of the skip connection to the normalized output. Equation \ref{equa:3} also demonstrates the skip connection output is adjustable by utilizing two learnable parameters, denoted as $W_{1}$ and $W_{2}$. They are applied in the Channel MLP layer showcasing this adjustment, where $W_{1}$ and $W_{2}$ are used to modify the output and shape it according to the desired characteristics or patterns. The activation function $\sigma(.)$ is applied to the output of the Channel MLP layer, introducing non-linearity and enhancing the expressive power of the ConvFormer block.

\subsection{Transformer Block}
The Transformer block incorporates the fundamental principles of the conventional Transformer. It consists of a traditional self-attention mechanism that generates an attention mask. By utilizing this self-attention mechanism, the model can focus on various segments of the input sequence and comprehend the interrelationships among its features. The attention feature is derived from the resemblance between the input tokens, which determines the significance of each token in the ultimate output. This mechanism facilitates the model in grasping distant connections and contextual details.


\begin{align}
    X_{i} = X_{i} + Self Attention(Norm(X_{i})) 
    \label{equa:4} \\
     X_{i} = X_{i} + \sigma(Norm(X_{i})W_{1})W_{2}
    \label{equa:5}
\end{align}

In Equation \ref{equa:4}, the self-attention mechanism, known as Self Attention, is introduced. This mechanism governs the computation of the output. Equation \ref{equa:5} represents the skip connection that is applied. The resulting output can be adjusted by utilizing two learnable parameters, namely $W_{1}$ and $W_{2}$, which originate from the Channel MLP layer.

\subsection{Multi-scale Upsampling Block}

The Multi-scale Upsampling block (MU) comprises two stages. The first is the Upsampling stage, where the input data is upsampled to a higher resolution. This process increases the spatial dimensions of the data, while the second stage is the Multi-scale Addition, where multiple scales or resolutions of the upsampled data are added together. This addition operation allows for incorporating information from different scales, enabling the model to capture fine-grained and high-level features. Combining information from multi-scale gives the model a more comprehensive understanding of the polyp foreground and background, thus making more accurate predictions.



\begin{align}
    X_{i} = Upsampling(X_{i})
    \label{equa:6}
    \\
    X_{i} = Conv_{3x3}(X_{i}) + Conv_{7x7}(X_{i}) 
    \label{equa:7} \\
    X_{i-2} = \sigma(X_{i-2} + X_{i}) 
    \label{equa:8}
\end{align}


Equation~\ref{equa:6} describes the Upsampling stage, in which the nearest interpolation is used. The $X_{i} \in R^{W \times H \times C}$ denotes the input tensor at stage $i$. Following this, the Multi-scale Addition is performed by extracting the input tensor through two convolution layers with kernel sizes of $3 \times 3$ and $7 \times 7$. Two output features are added to create the decoder's output features at stage ${i}$, as seen in Equation~\ref{equa:7}.

Finally, the Multi-scale Upsampling feature is added with the feature output at the decoder stage $i-2$ to form the new representation feature of the decoder. After that, the activation function $\sigma(.)$ is applied to the output, further enhancing the non-linear mapping and introducing the model's non-linearity; the full equation is described in Equation~\ref{equa:8}.

\section{Experiments}
\label{sec:Experiments}
\subsection{Dataset}

To conduct the fair comparison, we follow the merged dataset from the PraNet \cite{pranet} for training which includes 900 samples from Kvasir-SEG \cite{kvasir-seg} and 550 samples from CVC-ClinicDB \cite{clinicdb}. The remaining images of Kvasir-SEG \cite{kvasir-seg} and CVC-ClinicDB \cite{clinicdb} with three unseen datasets, including ColonDB \cite{colondb}, CVC-300 \cite{cvc300}  and ETIS \cite{etis} are used for benchmarking.

\subsection{Implementation Details and Evaluation Metrics}

All architectures are implemented using the Keras framework with TensorFlow as the backend. Input images are normalized to the range [-1, 1]. Adam optimization \cite{kingma2014adam} is utilized with an initial learning rate of 1e-4. Subsequently, the Cosine Annealing learning rate schedule stabilizes the training process. 

The experiments are conducted on a single NVIDIA Tesla A100 40GB GPU with a batch size of 128. Training the entire dataset takes approximately 6 hours. The model is trained for 158 epochs. The images are resized to 352 $\times$ 352 during the training and inference stages. Augmentation techniques such as Center Crop, Random Rotate, GridDistortion, CutOut \cite{cutout}, CutMix \cite{cutmix}, Horizontal and Vertical Flip are applied. 

The Jaccard loss function used to supervise our model in the training process can be formulated as follows: 

\begin{equation}
\centering
    Jaccard Loss(y,\hat y) = \alpha\times(1-\frac{\alpha+\sum_{c}^Cy_{c}\times\hat y_{c}}{\alpha +\sum_{c}^Cy_{c} + \hat y^{c} - y_{c}\times\hat y_{c}})
\centering 
\label{jac}
\end{equation}

The Jaccard Loss, which is shown in Equation \ref{jac}, is also called the Intersection over Union (IOU) metric \cite{jaacaard}. The true label is denoted as $y$, while the predicted label is represented as $\hat y$. Both labels are expressed as one-hot vectors, indicating the classes with a length equivalent to the number of classes, denoted as $C$. The $\alpha$ smoothing factor is set to 0.7 in our method. Note that we do not use deep supervision techniques to train our model.

We adopt three evaluation common metrics for quantitative evaluation: mean Dice (mDice), mean IoU (mIoU), and Mean Absolute Error (MAE) to evaluate our method's performance. The higher value is better for mDice as well as mIoU, and the lower is better for MAE.





\begin{table}[!h]
\begin{center}
\begin{tabular}{|c|c| c | c| c|} 
 \hline
  & Methods & $mIoU \uparrow$ & $mDice \uparrow$ & $MAE \downarrow$ \\ [0.5ex] 
 \hline
 \multirow{8}{*}{Kvasir} & UNet\cite{unet} & 0.756 & 0.821 & 0.055 \\

 & UNet++ \cite{unet++} &  0.743 & 0.820 & 0.048 \\

 & SFA \cite{sfa} & 0.611 & 0.723 & 0.075 \\

 & PraNet \cite{pranet}  & 0.840 & 0.898 & 0.030 \\

 & MSNet  \cite{MSNet} & \textbf{0.862} & \textbf{0.907} & \underline{0.028} \\

 & Hieu et al. \cite{rivf} &  0.835 & 0.891 & 0.029\\

 & PEFNet \cite{mmm2023} & 0.833 & 0.892 & 0.029 \\
 
 & \textbf{M$^2$UNet} & \underline{0.855} & \textbf{0.907} & \textbf{0.025} \\

 \hline

 \multirow{8}{*}{ClinicDB} & UNet \cite{unet} & 0.767 & 0.824 & 0.019 \\

 & UNet++ \cite{unet++} &  0.729 & 0.794 & 0.022 \\

 & SFA \cite{sfa} & 0.607 & 0.700 & 0.042 \\

 & PraNet \cite{pranet}  & 0.849 & 0.899 & \underline{0.009} \\

 & MSNet \cite{MSNet} & \textbf{0.879} & \textbf{0.921} & \textbf{0.008} \\

 & Hieu et al. \cite{rivf} &  0.787 & 0.844 & 0.019\\

 & PEFNet \cite{mmm2023} & 0.814 & 0.866 & 0.010 \\
 
 & \textbf{M$^2$UNet} & \underline{0.853} & \underline{0.901} & \textbf{0.008} \\
 \hline

 \multirow{8}{*}{CVC-300} & UNet \cite{unet} & 0.639 & 0.717 & 0.022 \\

 & UNet++ \cite{unet++} &  0.636 & 0.687 & 0.018 \\

 & SFA \cite{sfa} & 0.329 & 0.467 & 0.065 \\

 & PraNet \cite{pranet}  & 0.797 & 0.871 & 0.010 \\

 & MSNet \cite{MSNet} & 0.807 & 0.869 & 0.010 \\

 & Hieu et al. \cite{rivf} &  0.812 & 0.884 & \textbf{0.006}\\

 & PEFNet \cite{mmm2023} & 0.797 & 0.871 & 0.010 \\

 & \textbf{M$^2$UNet} & \textbf{0.819} & \textbf{0.890} & \underline{0.007} \\
 \hline

 \multirow{8}{*}{ColonDB} & UNet \cite{unet} & 0.449 & 0.519 & 0.061 \\

 & UNet++ \cite{unet++} &  0.410 & 0.483 & 0.064 \\

 & SFA \cite{sfa} & 0.347 & 0.469 & 0.094 \\

 & PraNet \cite{pranet}  & 0.640 & 0.712 & 0.043 \\

 & MSNet \cite{MSNet} & \underline{0.678} & \underline{0.755} & 0.041 \\

 & Hieu et al. \cite{rivf} &  0.626 & 0.694 & \underline{0.037}\\

 & PEFNet \cite{mmm2023} & 0.638 & 0.710 & \textbf{0.036} \\

 & \textbf{M$^2$UNet} & \textbf{0.684} & \textbf{0.767} & \textbf{0.036} \\
 \hline

 \multirow{8}{*}{ETIS} & UNet \cite{unet} & 0.343 & 0.406 & 0.036 \\

 & UNet++ \cite{unet++} &  0.344 & 0.401 & 0.035 \\

 & SFA \cite{sfa} & 0.217 & 0.297 & 0.109 \\

 & PraNet \cite{pranet}  & 0.567 & 0.628 & 0.031 \\

 & MSNet \cite{MSNet} & \textbf{0.664} & \textbf{0.719} & 
 \underline{0.020} \\

 & Hieu et al. \cite{rivf} &  0.589 & 0.655 & 0.037\\

 & PEFNet \cite{mmm2023} & 0.572 & 0.636 & \textbf{0.019} \\

 & \textbf{M$^2$UNet} & \underline{0.595} & \underline{0.670} & 0.024 \\
 \hline
 
\end{tabular}
\end{center}
\caption{Quantitative results with previous methods. \\The highest and second highest scores are shown in \\\textbf{bold} and \underline{underline}, respectively.}
\label{tab:quantitative}
\vspace{-6mm}
\end{table}

\subsection{Performance Comparisons}

To evaluate the effectiveness of our model, we compare M$^2$UNet with several methods, including UNet \cite{unet}, UNet++ \cite{unet++}, SFA \cite{sfa}, PraNet \cite{pranet}, MSNet \cite{MSNet}, Hieu \textit{et al.} \cite{rivf} and PEFNet \cite{mmm2023}. Since the setting datasets of Hieu \textit{et al.} \cite{rivf} and PEFNet \cite{mmm2023} are different, we retrain both methods on the same setting in PraNet \cite{pranet} for a fair comparison.

\vspace{2mm}
\subsubsection{Quantitative Result}

As shown in Table~\ref{tab:quantitative}, our M$^2$UNet attains superior performance in CVC-300 and ColonDB datasets, demonstrating our model's ability in the unseen domain. On the other three unseen datasets, \textit{i.e.} Kvasir, ClinicDB, and ETIS, the M$^2$UNet also obtains the second-highest score even though we do not utilize the deep supervision technique.





\vspace{2mm}

\subsubsection{Qualitative Result}

\begin{figure*}[!h]
    \centering
    \includegraphics[width=1\linewidth]{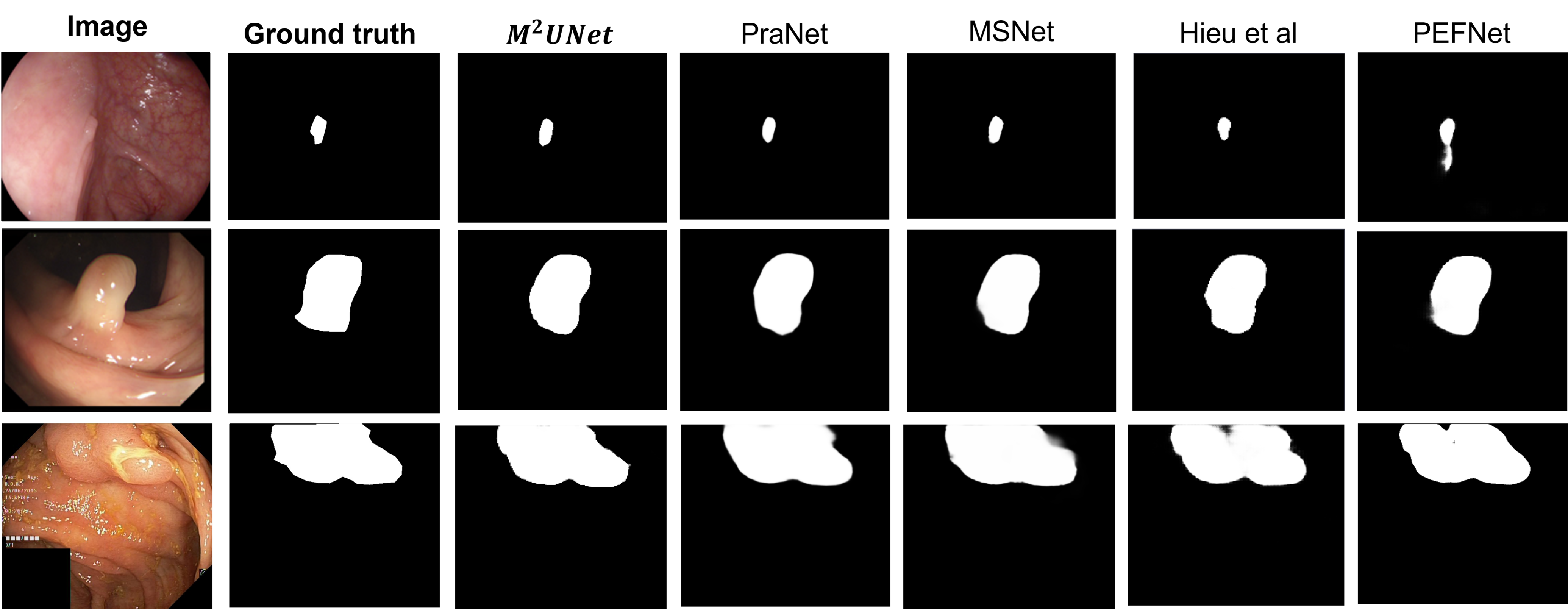}
    \caption{Qualitative results from various methods}
    \label{fig:vis}
\end{figure*}

In Figure~\ref{fig:vis}, we perform the qualitative visualization for several methods. It can be seen that our method can cover more accurate polyp regions. In general, our MU can better highlight the polyp regions based on the multi-scale information of different levels of the decoder.

\subsection{Ablation Study}
To evaluate the effectiveness of the MU module, we conduct the ablation study on the Kvasir dataset. As described above, the standard MU module includes two stages, Upsampling and Multi-scale Addition. We analyze the quantitative contribution of both Upsampling stage and MU module to the model performance, which is shown in Table~\ref{tab:ablation}.


According to the empirical findings, integrating the Metaformer backbone with UNet yields a dice score of 0.874. Including an additional Upsampling stage leads to a gradual improvement in the score, reaching 0.882. Introducing Multi-scale Upsampling further enhances the results slightly, raising the score to 0.891. 

By incorporating two Upsampling stages, the dice score improves to 0.899. Notably, when employing the complete pipeline, the score experiences a substantial boost, reaching 0.907.

\begin{table}[b!]
\begin{center}
\begin{tabular}{|c| c | c| c|} 
 \hline
 Model & $mIoU \uparrow$ & $mDice \uparrow$ & $MAE \downarrow$ \\ [0.5ex] 
 \hline
Baseline  & 0.819 & 0.874 & 0.029 \\
+1 Upsampling & 0.827 & 0.882 & 0.030 \\
+1 MU  & 0.833 & 0.891 & 0.029\\
+2 Upsampling & 0.843 & 0.899 & 0.028 \\
+2 MU & 0.855 & 0.907 & 0.025 \\
 
 \hline
\end{tabular}
\end{center}
\caption{Ablation study on the Kvasir-SEG dataset \cite{kvasir-seg}}
\label{tab:ablation}
\end{table}

The ablation study results demonstrate our assumption that capturing multi-level information of multiple decoder stages will ameliorate the model's ability to identify the polyp objects. By combining the Multi-scale Addition stage of our MU, the model can exploit multi-scale features of shallow stages to complement the necessary information further, thus attaining better performance.

\section{Conclusion}
\label{sec:Conclusion}

In this paper, we propose the MetaFormer baseline combined with UNet and our Multi-scale Upsampling block (MU) for the polyp segmentation task. Our M$^2$UNet is designed to solve the problem from existing methods, which is the locality of the standard convolution operations and the lack of attention to the multi-level information of the decoder's features. 

We believe that capturing multi-scale features of shallow decoder stages and combining them with the higher ones will endow our model to attain more meaningful information about the polyp regions. Extensive experiments and ablation results have demonstrated the improvement of our approach to previous methods. 

Although there are some limitations of the M$^2$UNet that need to be improved, this is a promising method for the polyp segmentation task. Future research should focus on leveraging the full potential of the decoder module for more precise and reliable polyp localization.

\section*{Acknowledgement} 
This research is funded by Vietnam National University Ho Chi Minh City (VNU-HCM) under grant number DS2020-42-01.

\bibliographystyle{IEEEtran}
\bibliography{myref.bib}

\end{document}